\documentclass[aps,pra,nofootinbib,preprint,amsmath,amssymb,floatfix]{revtex4}
\usepackage{color}
\usepackage{graphicx}

\begin{document}

\newcommand{\tc}{\textcolor}
\newcommand{\g}{blue}
\newcommand{\ve}{\varepsilon}
\title{Repulsive Casimir force}         

\author{Johan S. H\o ye$^1$ and  Iver Brevik$^2$  }      
\affiliation{$^1$Department of Physics, Norwegian University of Science and Technology, N-7491 Trondheim, Norway}

\affiliation{$^2$Department of Energy and Process Engineering, Norwegian University of Science and Technology, N-7491 Trondheim, Norway}
\date{\today}          

\begin{abstract}

The Casimir force between two parallel thick plates, one perfectly dielectric, the other purely magnetic, has been calculated long ago by Boyer [T. H. Boyer, Phys. Rev. A {\bf 9}, 2078 (1974)]. Its most characteristic property is that it is repulsive. The problem is actually delicate and counterintuitive.
In the present paper we analyze the problem by first considering the simple harmonic oscillator model
introduced by us earlier [J. S. H{\o}ye {\it et al.}, Phys. Rev. E {\bf 67}, 056116 (2003); Phys. Rev. A {\bf 94}, 032113 (2016)]. Extension of this model shows how the repulsive behavior can be understood on a microscopic basis, due to the duality between canonical and mechanical momenta in presence of the electromagnetic vector potential. This duality  corresponds to the TM and TE modes in electrodynamics. We analyze the generalized Boyer  case where the permittivities and permeabilities of the parallel plates are arbitrary. In this respect we first find the induced interaction between a pair of particles with given electric and magnetic polarizabilities and then find it for a pair of parallel plates. The method used for our evaluations is the statistical mechanical one that we have introduced and applied earlier. Whether the pair of particles or plates attract or repel each other depends on their polarizabilities or permittivities and permeabilities respectively. For equal particles or equal plates there is always attraction.

\end{abstract}
\maketitle

\bigskip
\section{Introduction}
\label{secintro}
The Casimir force between two equal parallel isotropic nonmagnetic materials separated by a vacuum (air) gap $a$ is under usual circumstances known to be attractive  (for reviews on the Casimir effect. cf., for instance, Ref.~\cite{bordag09,milton01,parsegian06}). It turns out to be possible, however, in some cases to  make the Casimir force switch sign so that it becomes repulsive. Typically, this occurs if  a special  liquid like bromobenzene is immersed between  gold and silica surfaces as shown by Munday {\it et al.} \cite{munday09}. Then Casimir repulsion takes place at distances above approximately 5 nm since the dielectric constant of bromobenzene lies between those of gold and silica for imaginary frequencies below a certain value \cite{bostrom12,wennerstrom99,ninham98}.

The analog case of two equal purely magnetic plates behaves in the same way; the Casimir force is under usual circumstances attractive.

The case of two unequal plates - one purely dielectric and the other purely magnetic - is however different in the sense that the Casimir force becomes {\it repulsive}. This peculiar effect has been known for several years. A well known reference in this area is the paper of Boyer \cite{boyer74} (the present problem is sometimes called the Boyer problem). Based upon earlier investigations of Feinberg and Sucher \cite{feinberg68} on the van der Waals force between two electrically and magnetically particles, he calculated the repulsive force between a conducting and a permeable plate in the limit $\varepsilon \rightarrow \infty, \mu \rightarrow \infty$. Boyer's formalism was based upon so-called random electrodynamics.

We ought to point here that this change of force direction is actually quite nontrivial. Consider for definiteness the left plate with boundary  $z=0$ to be purely dielectric, with finite permittivity $\varepsilon >1$, while the right plate with boundary $z=a$ is purely magnetic with permeability $\mu > 1$. In the boundary region of the dielectric plate around $z=0$ the only electromagnetic volume force density is $-\frac{1}{2}\varepsilon_0 E^2\nabla \varepsilon$. Here the $E$ is the magnitude of the varying electric field from the outside to the inside of the dielectric boundary. This should be expected to give a surface force density acting in the positive $z$ direction, i.e.  toward the vacuum region, in accordance with
the general property of classical electrodynamics saying that the surface force acts in the direction of the optically thinner medium. In our case the force direction is however reversed. Taken literally, this has to be interpreted to imply that the square $E^2$ becomes negative, thus a counterintuitive result whose physical significance is not obvious.

We mention that the problem has been considered in the literature from a more wide viewpoint also, not restricted to the idealized case of Boyer where one plate was electric and the other magnetic. Some papers dealing with these topics are  Refs. \cite{richmond78,kenneth02,inui11,pappakrishnan14,sinha18}. Most of them, such as Refs.~\cite{kenneth02,inui11,pappakrishnan14}, analyze Casimir repulsive forces in magnetodielectric configurations, while the very recent Ref.~\cite{sinha18} analyzes the repulsive force on a magnetic point particle near a surface. In general, whether the force is found to be repulsive or attractive, depends on the relative strengths of $\varepsilon_i$ and $\mu_1$ ($i=1,2$). The delicate point mentioned above - that a Casimir calculation leads to an effective square $E^2 <0$ while always  $E^2 \ge 0$ for $E$ real - will not be investigated further in this work.
In the present paper our purpose is instead  to analyze the Boyer problem in a generalized sense, i.e. take the polarizabilities of the two particles or the permittivities and permeabilities to be general.

We start  from a microscopic harmonic oscillator model where two oscillators interact via a third one. We have employed this model before \cite{hoye03}. The model was later extended to cover the case where the third oscillator describes a set of oscillators \cite{hoye16}.
 The two oscillators of the model correspond to two polarizable particles that interact via the harmonic oscillator modes of the electromagnetic field. For the latter there are two types of waves that are clearly noticed when a pair of parallel polarizable plates are considered. They are the TM (transverse magnetic) and TE (transverse electric) waves or modes. The former is the one most easily understood in view of the induced attractive interaction. Also the induced interaction increases in magnitude with temperature to reach the classical limit. For the TE mode, however, the situation is less obvious as the force decreases to zero in the classical limit since its Matsubara frequency $\zeta=0$ does not contribute. To cope with this situation interaction terms similar to the interaction of a particle with the electromagnetic vector potential was used.

To obtain the situation corresponding to a dielectric plate interacting with a magnetic one, we find that our model can be extended in a straightforward way to encompass such a  situation. During computations two quantities $D_1$ and $D_2$ appear  (defined in Eq.~(\ref{18}) below), where their product $D_1 D_2>0$ determines an induced attractive force. Now one notes that for the analog of the TM case both quantities are positive while for the TE case both are negative by which  $D_1 D_2>0$ remains.

We will show how  the model can be extended such that $D_1$ and $D_2$ have opposite signs. This will be the analog of a pair of dielectric and magnetic plates that repel each other.

In Sec.~\ref{sec2} we consider the harmonic oscillator model, showing how the transition from attractive to repulsive force between dipoles can be traced back to the fundamental distinction between mechanical momentum and canonical momentum in classical electrodynamics. In Sec.~\ref{sec3} we derive the induced interaction between a pair of particles where there are couplings between both electric and magnetic dipole moments. In Sec.~\ref{sec4} we generalize the statistical mechanical derivations of Ref.~\cite{hoye98} to two dielectric half-planes with different dielectric constants. From this the well-known Lifshitz result is recovered. Finally, in Sec.~\ref{sec5} this is generalized further to obtain the Casimir force with magnetic properties included. Section VI  summarizes the obtained results.

\section{Harmonic Oscillator model for repulsive Casimir force}
\label{sec2}

Now consider the harmonic oscillator model treated in  Refs.~\cite{hoye03} and \cite{hoye16}. We will here only sketch the previous derivations while more details can be found in the mentioned references. Let the three oscillators have eigenfrequencies $\omega_i$ ($i=1,2,3$). Their classical partition function is then proportional to the inverse of $\sqrt{Q}$ where
\begin{equation}
Q=a_1 a_2 a_3, \quad a_i=\omega_i^2 \quad (i=1,2,3).
\label{10}
\end{equation}
For simplicity all three oscillators of the model are one-dimensional.

By quantization using the path integral method \cite{hoye81,brevik88} the classical system turns out to be split into a set of classical harmonic oscillator systems where the Matsubara frequencies are added to the eigenfrequencies. This replaces expression (\ref{10}) by
\begin{equation}
Q=A_1 A_2 A_3, \quad A_i=a_i+\zeta^2 = \omega_i^2+\zeta^2
\label{11}
\end{equation}
where $\zeta=i\omega$ (the convention $\zeta=-i\omega$ may also be used).

When the oscillators interact (via bilinear terms) the $Q$ can be expressed through a determinant. With interaction parameter $c$ for the interactions via oscillator 3, one obtains the result (9) of Ref.~\cite{hoye16} (or (4.3a) of \cite{hoye03})
\begin{eqnarray}
Q&=&\left|
\begin{array}{ccc}
A_1 & 0 & c\\
0 & A_2 & c\\
c & c & A_3\\
\end{array}
\right|
=A_1 A_2 A_3 -c^2(A_1+A_2)=A_1 A_2 A_3(1-D_1-D_2)
\nonumber\\
&=&A_1 A_2 A_3(1-D_1)(1-D_2)\left(1-\frac{D_1 D_2}{(1-D_1)(1-D_2)}\right),
\label{12}
\end{eqnarray}
\begin{equation}
D_j=\frac{c^2}{A_j A_3}\quad (j=1,2).
\label{13}
\end{equation}
Here the $A_j(1-D_j)$ terms represent each of the two oscillators and their radiation reaction with the third oscillator while the last factor corresponds to the Casimir free energy. With $D_j$ given by (\ref{13}) this factor is larger than zero which means that the induced force between the two oscillators is attractive. This situation is the analog of the TM mode.

The second situation is the analog of the TE mode. To analyze this we may start from the situation where oscillator 3 interacts  with the momenta of oscillators 1 and 2.
Then the interaction with the third oscillator is like the interaction via the electromagnetic vector field. With this field the  Hamiltonian for a particle $j$ is $({\bf p}_j-(e/c){\bf A})^2/2m_j$, where here $c$ is the velocity of light, ${\bf p}_j$ the canonical momentum, and ${\bf A}={\bf A}({\bf r}_j)$ the vector potential (in Gaussian units). The mechanical momentum is ${\bf p}_{Mj}={\bf p}_j-(e/c){\bf A}$. In our model the analogous interaction can be written as $a_j(p_j-(c/a_j)x_3)^2$ ($j=1,2$), where again $c$ is the coupling parameter and the coordinate $x_3$ corresponds formally to the vector potential ${\bf A}$.

 Now for harmonic oscillators the roles of the momenta and coordinates can be exchanged to obtain the energy term
\begin{equation}
a_j\left(x_j-\frac{c}{a_j}x_3\right)^2.
\label{14}
\end{equation}
With this the $A_3$ term in the determinant (\ref{12}) is changed into
\begin{equation}
A_3\rightarrow A_3+\frac{c^2}{a_1}+\frac{c^2}{a_2}.
\label{15}
\end{equation}
 The rows $j=1,2$ of the determinant can be multiplied with $c/a_1$ and $c/a_2$ respectively to be subtracted from row 3. Then  rows 1 and 2 can be multiplied with $\zeta/\sqrt{a_j}$, and columns 1 and 2 can be divided with the same factors. These operations do not change the determinant. We obtain
\begin{eqnarray}
Q&=&\left|
\begin{array}{ccc}
A_1 & 0 & c\\
0 & A_2 & c\\
cq_1 & cq_2 & A_3\\
\end{array}
\right|
=\left|
\begin{array}{ccc}
A_1 & 0 & \zeta c/\sqrt{a_1}\\
0 & A_2 & \zeta c/\sqrt{a_2}\\
-\zeta c/\sqrt{a_1} &-\zeta c/\sqrt{a_2}  & A_3\\
\end{array}
\right|,
\label{16}
\end{eqnarray}
\begin{equation}
q_j=1-\frac{A_j}{a_j}=-\frac{\zeta^2}{a_j}, \quad(j=1,2).
\label{17}
\end{equation}
 With determinant (\ref{16}) the $Q$ when evaluated can still be written as result (\ref{12}) but now with
\begin{equation}
D_j=-\frac{\zeta^2 c^2}{a_j A_j A_3}<0.
\label{18}
\end{equation}
Again the induced force will be attractive since $D_1 D_2>0$, but this time both factors are negative. However, in the classical limit where only $\zeta=0$ contributes this force will vanish as also is the case for the TE mode.

From the above it is now easily seen that the induced force can be repulsive too. This will be the case if only one of the energies has form (\ref{14}), say for $j=2$ but not for $j=1$. Then instead of (\ref{15}) the change of $A_3$ will be
\begin{equation}
A_3\rightarrow A_3+\frac{c^2}{a_2}.
\label{19}
\end{equation}
The determinant (\ref{16}) will be modified to
\begin{eqnarray}
Q=\left|
\begin{array}{ccc}
A_1 & 0 &  c\\
0 & A_2 & \zeta c/\sqrt{a_2}\\
c &-\zeta c/\sqrt{a_2}  & A_3\\
\end{array}
\right|,
\label{17}
\end{eqnarray}
by which relation (\ref{18}) changes into
\begin{equation}
D_1=\frac{c^2}{A_1 A_3}>0 \quad \mbox{and} \quad D_2=-\frac{\zeta^2 c^2}{a_j A_j A_3}<0.
\label{21}
\end{equation}
With this $D_1 D_2<0$ by which the induced force will be repulsive. This force will be the analog of the repulsive force between a dielectric medium and a magnetic one.

\section{Induced interaction between a pair of particles with both dielectric and magnetic polarization}
\label{sec3}

In the static case electric and magnetic properties separate. However, for non-zero frequencies there is a coupling between electric and magnetic dipole moments. The fields are determined by Maxwell's equations which in standard notation can be written (Gaussian units)
\begin{eqnarray}
\nonumber
\nabla {\bf D}&=&0,\quad \nabla \times{\bf E}=-\frac{\partial{\bf B}}{c\,\partial t},\quad {\bf D}={\bf E}+4\pi{\bf P}=\varepsilon {\bf E},\\
\nabla {\bf B}&=&0,\quad \nabla \times{\bf H}=\frac{\partial{\bf D}}{c\,\partial t},\quad {\bf B}={\bf H}+4\pi{\bf M}=\mu {\bf H}.
\label{30}
\end{eqnarray}
The equations can be Fourier transformed by which
\begin{equation}
\nabla\rightarrow-i{\bf k}\quad\mbox{and}\quad\frac{\partial}{c\,\partial t}\rightarrow i\frac{\omega}{c}=\zeta.
\label{31}
\end{equation}
(The signs of Fourier transforms depend upon the convention chosen.) Note, here and below definition (\ref{31}) is used for $\zeta$.

For a given polarization one finds the solution \cite{hoye82}
\begin{eqnarray}
\nonumber
{\bf E}&=&-\frac{4\pi}{3}\left\{\frac{k^2}{k^2+\zeta^2}[3(\hat{\bf k}\cdot{\bf P})\hat{\bf k} -{\bf P}]+\left[\frac{2\zeta^2}{k^2+\zeta^2}+1\right]{\bf P}\right\}\\
&=&-\frac{4\pi}{3}\left\{\frac{k^2}{k^2+\zeta^2}3(\hat{\bf k}\cdot{\bf P})\hat{\bf k} +\frac{3\zeta^2}{k^2+\zeta^2}{\bf P}\right\}
\label{32}
\end{eqnarray}
where the hat is used for unit vectors. For simplicity the same vector notation is used both in $r$-space and $k$-space as the argument used will specify. In $r$-space (\ref{32}) becomes \cite{hoye82}
\begin{equation}
{\bf E}={\bf E_1}=\frac{e^{-x}}{r^3}\left\{\left(1+x+\frac{1}{3}x^2\right)[3(\hat{\bf r}\cdot{\bf s}_1)\hat{\bf r}-{\bf s}_1]-\frac{2}{3}x^2{\bf s}_1\right\}-\frac{4\pi}{3}\delta({\bf r}){\bf s}_1
\label{33}
\end{equation}
where $x=\zeta r$. Here the polarization used is a point dipole
\begin{equation}
{\bf P}={\bf P}({\bf r})={\bf s}_1\delta({\bf r}), \quad ({\bf P}={\bf P}({\bf k})={\bf s}_1).
\label{34}
\end{equation}
The interaction with a dipole moment ${\bf s}_2$ at position ${\bf r}$ is then $-{\bf E}_1{\bf s}_2$ which also is proportional to the correlation function for the two dipoles ${\bf s}_1$ and ${\bf s}_2$. Thus for (large) $r$ the induced free energy is proportional to $\langle({\bf E}_1{\bf s}_2)^2\rangle$ where the thermal average is over the fluctuating dipole moments for each value of $\zeta$. With $\langle({\bf a}\cdot\hat{\bf s})({\bf b}\cdot\hat{\bf s})\rangle={\bf a}\cdot{\bf b}/3$ one obtains
\begin{equation}
\langle({\bf E}_1\cdot{\bf s}_2)^2\rangle=\frac{1}{3}\langle s_1^2\rangle\langle s_2^2\rangle\frac{1}{r^6}L_{EE}(x),
\label{36}
\end{equation}
\begin{equation}
L_{EE}(x)=e^{-2x}\left[2\left(1+x+\frac{1}{3}x^2\right)^2+\left(\frac{2}{3}x^2\right)^2\right].
\label{37}
\end{equation}

Now the electric polarizability is given by ($i=1,2$)
\begin{equation}
\alpha_{iE}=\alpha_{iE}(\zeta))\propto \langle s_i^2\rangle.
\label{38}
\end{equation}
With other details that follow from the derivations in Ref.~\cite{brevik88}, the induced free energy becomes Eq.~(5.15) of the reference
\begin{equation}
F_{EE}=-\frac{3}{2\beta r^6}\sum_{n=-\infty}^\infty L_{EE}(x)\alpha_{1E}\alpha_{2E}
\label{39}
\end{equation}
with $\beta=1/(k_B T)$ where $k_B$ is Boltzmann's constant. (The factor 2 in the denominator is missing in the reference.) The sum is over the Matsubara frequencies
\begin{equation}
x=\zeta r=\frac{Kr}{\hbar c},\quad K=\frac{2\pi n}{\beta}\quad\mbox{with $n$ integer}.
\label{40}
\end{equation}
At $T=0$ the sum can be replaced by an integral with
\begin{equation}
dx=r\,d\zeta=\frac{2\pi r}{\hbar c\beta}\,dn\quad\mbox{or}\quad \sum_{n=-\infty}^\infty\rightarrow2\frac{\hbar c\beta}{2\pi r}\int_0^\infty dx.
\label{41}
\end{equation}
With constant (i.e.~independent of $\zeta$) polarizabilities this further gives
\begin{equation}
F_{EE}=-\frac{3\hbar c \alpha_{1E}\alpha_{2E}}{2\pi r^7}I_{EE}
\label{42}
\end{equation}
\begin{equation}
I_{EE}=\int\limits_0^\infty L_{EE}(x)\,dx=\frac{23}{6}
\label{43}
\end{equation}
which is the well established result (5.16) of Ref.~\cite{brevik88}.

Another situation is the induced interaction between the magnetic moments of the two particles. But due to the symmetry of Maxwell's equations (\ref{30}) this will be precisely like results (\ref{39}) and (\ref{42}) with the subscript $_E$ replaced by $_H$.

What remains is the induced interaction between electric and magnetic moments. To do so we need to obtain  the $L_{EH}$ function to replace $L_{EE}$ and replace an electric polarizability with a magnetic one. So consider the electric field created by a polarization. For non-zero frequencies it also creates a magnetic field. From Eqs.~(\ref{30}) and (\ref{31}) one has
\begin{equation}
-i({\bf k}\times{\bf E})=-i\frac{\omega}{c}{\bf B}=-\zeta {\bf H},\quad {\bf B}={\bf H}\,\,({\bf M}=0).
\label{44}
\end{equation}
With the last equality of Eq.~(\ref{32}) this simplifies to
\begin{equation}
{\bf H}=\frac{i}{\zeta}\left({\bf k}\times\left(\frac{-4\pi\zeta^2}{k^2+\zeta^2}{\bf P}\right)\right)
\label{45}
\end{equation}
(since $\hat{\bf k}\times\hat{\bf k}=0$). From this follows ($\nabla\times(\psi{\bf P})=(\nabla\psi)\times{\bf P}+\psi\nabla\times{\bf P}$)
\begin{equation}
{\bf H}=-\frac{1}{\zeta}\nabla\times\left(\frac{-\zeta^2e^{-\zeta r}}{r}{\bf P}\right)=\tc{red}{-}\zeta(1+\zeta r)\frac{e^{-\zeta r}}{r^2}(\hat{\bf r}\times{\bf P}).
\label{46}
\end{equation}
So the interaction of an electric dipole moment ${\bf s}_1$ with a magnetic one ${\bf s}_2$ is
\begin{equation}
-{\bf H}_1{\bf s}_2=-\zeta(1+\zeta r)\frac{e^{-\zeta r}}{r^2}(\hat{\bf r}\times{\bf s}_1)\cdot{\bf s}_2.
\label{47}
\end{equation}

The average of interest is the square of (\ref{47}), but now with minus sign when compared with expression \eqref{36}. This change of sign is due to a property of Fourier transforms. For two functions $f(x)$ and $g(x)$ and their Fourier transforms $\tilde f(k)$ and $\tilde g(k)$ one har the relation
\begin{equation}
\int f(x) g(x)\,dx = \frac{1}{2\pi}\int \tilde f(k) \tilde g(-k)\,dk.
\label{48a}
\end{equation}
This relation is similar for situations when $x$ and/or $k$ have discrete values too. In the present case the $x$ corresponds to imaginary time with  $\zeta$ the Fourier variable. In the free energy expression \eqref{39} the sum over $\zeta$corresponds to the $k$ integration in relation \eqref{48a}. The transform \eqref{47} changes sign when $\zeta\rightarrow-\zeta$ (as its $\zeta r=|\zeta|r$).

In standard notation one can write $({\bf r}\times{\bf s}_1){\bf s}_2=\varepsilon_{ijk}x_i s_{1j}s_{2k}$ to obtain the average
\begin{equation}
\langle[({\bf r}\times{\bf s}_1){\bf s}_2]^2\rangle=\varepsilon_{ijk}\varepsilon_{npq}x_i x_n\langle s_{1j}s_{1p}\rangle\langle s_{2k}s_{2q}\rangle=\frac{2}{9}r^2\langle s_1^2\rangle\langle s_2^2\rangle.
\label{49}
\end{equation}
Thus
\begin{equation}
\langle({\bf H}_1\cdot{\bf s}_2)^2\rangle=\frac{1}{3}\langle s_1^2\rangle\langle s_2^2\rangle\frac{1}{r^6}L_{EH}(x)
\label{50}
\end{equation}
\begin{equation}
L_{EH}(x)=\frac{2}{3}e^{-2x}[x(1+x)]^2.
\label{51}
\end{equation}

Like Eq.~(\ref{39}), but now with opposite sign, the induced free energy expression   becomes
\begin{equation}
F_{EH}=\frac{3}{2\beta r^6}\sum_{n=-\infty}^\infty L_{EH}(x)\alpha_{1E}\alpha_{2H}
\label{52}
\end{equation}
plus the contribution $F_{HE}$ which will be the same except the product of polarizabilities that is modified into $\alpha_{1H}\alpha_{2E}$. With constant polarizabilities this similar to (\ref{42}) gives
\begin{equation}
F_{EH}=\frac{3\hbar c \alpha_{1E}\alpha_{2H}}{2\pi r^7}I_{EH}
\label{53}
\end{equation}
\begin{equation}
I_{EH}=\int\limits_0^\infty L_{EH}(x)\,dx=\frac{7}{6}
\label{54}
\end{equation}
This part of the free energy gives a repulsive force in agreement with earlier derivations \cite{boyer74}.

\section{Induced interaction between a pair of dielectric half-planes}
\label{sec4}

For a pair of half-planes one can generalize the statistical mechanical derivations of H{\o}ye and Brevik where two dielectric half-planes with the same dielectric constant were considered \cite{hoye98}. First we will reconsider this situation and modify it somewhat to include extension to half-planes with different dielectric constants. With this modification the method becomes more suitable for the extension to include different magnetic permeabilities too. Since equations of Ref.~\cite{hoye98} will be used repeatedly, we will designate them with the numeral I when referred to.

The half-planes with separation $a$ are parallel to the $xy$ plane. With this the dipole-dipole interaction can be Fourier transformed along the $x$ and $y$ directions to obtain Eq.~(I6.4) for the radiating interaction in vacuum between electric dipoles ${\bf s}_1$ and ${\bf s}_2$ ($z\neq0$)
\begin{equation}
\hat\phi(12)=-2\pi s_1 s_2\frac{e^{-q|z|}}{q}
[({\bf h}\cdot\hat{\bf s}_1)({\bf h}\cdot\hat{\bf s}_2)-\zeta^2\hat{\bf s}_1\cdot\hat{\bf s}_2]
\label{60}
\end{equation}
\begin{equation}
{\bf h}=\{ik_x, ik_y,\pm q\}, \quad q^2=k_\perp^2+\zeta^2, \quad k_\perp^2=k_x^2+k_y^2,
\label{61}
\end{equation}
Vectors with hats mean unit vectors.
Without loss of generality the ${\bf k}$ can be directed along the $x$ axis by which ${\bf h}=\{ik_\perp, 0,\pm q\}$.
With Eq.~(I6.5) the interaction can be split in two parts
\begin{equation}
\hat\phi(12)=-2\pi s_1 s_2\frac{e^{-q|z|}}{q}(H_1+H_2)
\label{63}
\end{equation}
\begin{equation}
H_1=({\bf h}\cdot\hat{\bf s}_1)({\bf h}\cdot\hat{\bf s}_2)-\zeta^2\hat{\bf s}_1^{||}\cdot\hat{\bf s}_2^{||}, \quad H_2=-\zeta^2\hat{\bf s}_1^\perp\cdot\hat{\bf s}_2^\perp
\label{64}
\end{equation}
where ${\bf s}^{||}$ is the component of ${\bf s}$ in the ${\bf h}$ plane while ${\bf s}^\perp$ is transverse to it. These two cases are the separation into the TM (transverse magnetic) and TE (transverse electric) modes respectively.

For a medium with dielectric constant $\ve$ and magnetic permeability $\mu$ the  interaction (\ref{63}) is  modified into
\begin{equation}
\hat\phi_\ve(12)=-2\pi s_1 s_2\frac{e^{-q_\ve|z|}}{q_\ve}\left(\frac{1}{\ve}H_{\ve1}+\mu H_2\right)
\label{66}
\end{equation}
\begin{equation}
H_{\varepsilon1}=({\bf h}_\varepsilon\cdot\hat{\bf s}_1)({\bf h}_\varepsilon\cdot\hat{\bf s}_2)-\varepsilon\mu\zeta^2\hat{\bf s}_1^{||}\cdot\hat{\bf s}_2^{||}
\label{65}
\end{equation}
\begin{equation}
{\bf h}_\varepsilon=\{ik_\perp, 0, \pm q_\varepsilon\}, \quad q_\varepsilon^2=k_\perp^2+\varepsilon\mu\zeta^2.
\label{62}
\end{equation}
while $H_2$ is unchanged.

The interaction is due to the electric field from ${\bf s}_1$ that acts upon ${\bf s}_2$, or vice versa. Therefore it should be symmetric in the two dipole moments also when the dielectric constants are different. Thus $H_{\ve1}$ must be modified. With use of (\ref{62}) one finds Eq.~(I6.10) which can be written as
\begin{equation}
H_{\ve1}={\bf G}_{\ve1}\cdot\hat{\bf s}_2=g_{\ve1} g_{\ve2}, \quad
{\bf G}_{\ve1}= g_{\ve1}\frac{q_\ve}{k_\perp}{\bf u}_{\ve\pm}, \quad {\bf u}_{\ve\pm}=\left\{ik_\perp, 0, \pm\frac{k_\perp^2}{q_\ve}\right\}
\label{67a}
\end{equation}
\begin{equation}
g_{\ve i}=|{\bf h}_\ve\times\hat{\bf s}_i|=\frac{q_\ve}{k_\perp}\left(ik_\perp\hat s_{\perp i}\pm\frac{k_\perp^2}{q_\ve}\hat s_{||i}\right)=\left(iq_\ve\hat s_{\perp i}\pm k_\perp\hat s_{||i}\right), \quad (i=1,2)
\label{67b}
\end{equation}
Here the $\hat s_\perp$ and $\hat s_{||}$ are the components of $\hat {\bf s}^{||}$ along the direction of ${\bf k}_\perp$ (in the $xy$ plane) and the $z$ axis respectively. The advantage of this form is that it can be extended to half-planes with different dielectric constants with energy symmetric in ${\hat s}_1$ and ${\hat s}_2$. Thus Eq.~(\ref{67a}) can be modified to
\begin{equation}
H_{\ve1}={\bf G}_{\ve1}\cdot\hat{\bf s}_2=g_{{\ve_1}1} g_{{\ve_2}2}, \quad
{\bf G}_{\ve1}= g_{{\ve_1}1}\frac{q_{\ve_2}}{k_\perp}{\bf u}_{{\ve_2}\pm}
\label{68}
\end{equation}
For the interaction in vacuum ($\ve, \mu\rightarrow1$ one has similar expressions where $q_\ve\rightarrow q$. (Note that with vector ${\bf u}_{{\ve_2}\pm}$ the $\nabla{\bf D}=0$ of Eq.~(\ref{30}) is fulfilled.) The resulting Green function (electric interaction) for the two half-planes can now still be written in the form of Eq.~(I6.20)
\begin{equation}
\hat\phi_g(12)=J_\ve\left(D_E^{||}H_{\ve1}+D_E^\perp H_2\right),
\quad J_\ve=-2\pi s_1 s_2\frac{e^{-q_{\ve_1}u_1-q_{\ve_2}(u_2+a)}}{q_{\ve_1}}
\label{69}
\end{equation}
where $u_1=-z_1>0$ and $u_2=z_2-a>0$. The sought coefficients $D_E^{||}$ and $D_E^\perp$ will make the resulting expression fully symmetric. (Within one medium this becomes interaction (\ref{66}).)

With several parallel layers the ${\bf u}_{\ve\pm}$ and some other factors will change along with the change of the dielectric constant for various layers. But the dielectric constant for the half-plane $z<0$ stays fixed at $\varepsilon_1$. Thus for two half-planes the factors not containing $\varepsilon_2$ or $\mu_2$ can be separated out to be put into a quantity $L$ when determining the electric field through the layers. With this the electric field for the TM mode ($\sim{\bf G}_{\ve1}\sim e^{\mp q_{\varepsilon_1}z}q_{\varepsilon_1}{\bf u}_{\varepsilon_1\pm}$) will be a slight modification of Eq.~(I6.14)
\begin{equation}
{\bf E}/L=\left\{
\begin{array}{ll}
q_{\varepsilon_1}\left(\frac{1}{\ve_1}e^{-q_{\varepsilon_1}z}q_{\varepsilon_1}{\bf u}_{\varepsilon_1+}+Be^{q_{\varepsilon_1}z}{\bf u}_{\varepsilon_1-}\right),\quad & z_0<z<0\\
q(Ce^{-qz}{\bf u}_+ +C_1 e^{qz}{\bf u}_-),  \quad  & 0<z<a\\
q_{\varepsilon_2}De^{-q_{\varepsilon_2} z}{\bf u}_{\varepsilon_2+},\quad & a<z.
\end{array}
\right.
\label{70}
\end{equation}
where $B$, $C$, $C_1$, and $D$ are coefficients to be determined (relative to the $1/\ve_1$ term of \eqref{64}).

One condition to determine the coefficients is that the component of $\bf E$ along the $xy$ plane is continuous. However, for the other condition we here find it more convenient to turn to the magnetic field (instead of continuous ${\bf D}$ field normal to the interfaces) due to later extension to magnetic media. Then from Eq.~(\ref{30}) we will find Eq.~(I6.17)
\begin{equation}
{\bf h}_\varepsilon\times{\bf E}=\zeta{\bf B}=\zeta\mu {\bf H}.
\label{71}
\end{equation}
With ${\bf E}/L=e^{\mp q_\ve z} q_{\varepsilon}{\bf u}_{\varepsilon\pm}$  one finds (for $\varepsilon=1$ and $\varepsilon=\varepsilon_i$, $i=1,2$)
\begin{equation}
\zeta{\bf B}/L=\{ik_\perp, 0, \pm q_\varepsilon\}\times\{ik_\perp, 0, \pm\frac{k_\perp^2}{q_\varepsilon}\}q_\varepsilon e^{\mp q_\ve z}=\pm ik_\perp(q_\varepsilon^2-k_\perp^2)e^{\mp q_\ve z}\tc{red}{\hat{\bf e}_y}=\pm ik_\perp\varepsilon\mu \zeta^2e^{\mp q_\ve z}\hat{\bf e}_y
\label{72}
\end{equation}
where $\hat{\bf e}_y$ is the unit vector in the $y$ direction.

The components of ${\bf E}$ and ${\bf H}$ parallel to the the interfaces are both continuous by which one gets the equations
\begin{eqnarray}
\nonumber
q_\ve\left(\frac{1}{\ve_1}+B\right)&=&q\left( C+C_1\right)\\
\nonumber
\ve_1\left(\frac{1}{\ve_1}-B\right)&=& C-C_1\\
\label{73}
q(Ce^{-qa}+C_1e^{qa})&=&q_{\ve_2} De^{-q_{\ve_2}a}\\
\nonumber
Ce^{-qa}+C_1e^{qa}&=&\ve_2 De^{-q_{\ve_2}a}
\end{eqnarray}
The two last equations can be solved for $C$ and $C_1$ to be inserted in the equation that results from the elimination of $B$. This extends solution (I6.15) for $D$ to be
\begin{equation}
D=D_E^{||}=\frac{4\kappa_1 e^{(q_{\ve_2}-q)a}}{(\ve_1+\kappa_1)(\ve_2+\kappa_2)[1-A_n e^{-2qa}]},
\label{74a}
\end{equation}
\begin{equation}
A_n=\frac{(\ve_1-\kappa_1)(\ve_2-\kappa_2)}{(\ve_1+\kappa_1)(\ve_2+\kappa_2)}, \quad \kappa_i=\frac{q_{\ve_i}}{q}\quad (i=1,2).
\label{74b}
\end{equation}
The subscript $n=1,2,3,\cdots$ indicate the Matsubara frequencies (\ref{40}) $\zeta=\zeta_n=K/(\hbar c)$, $K=2\pi n/\beta$.

For the TE mode the electric field lies along the $y$ axis transverse to the ${\bf h}$ plane. Thus this component of the field can be written like Eq.~(I6.18)
\begin{equation}
E/L=\left\{
\begin{array}{ll}
\mu_1 e^{-q_{\varepsilon_1}z}+Be^{q_{\varepsilon_1}z},\quad & z_0<z<0\\
Ce^{-qz} +C_1 e^{qz},  \quad  & 0<z<a\\
De^{-q_{\varepsilon_2}z},\quad & a<z.
\end{array}
\right.
\label{76}
\end{equation}
where again $L$ is a quantity independent of $\ve_2$ and $\mu_2$.
The corresponding magnetic field will have components along both the $x$ and $z$ directions
\begin{equation}
\zeta\mu{\bf H}/L=\zeta{\bf B}/L={\bf h}_\varepsilon\times{\bf E}/L= \{ik_\perp, 0, \pm q_\varepsilon\}\times\{0, 1, 0\}e^{\mp q_\ve z}=(\mp q_\ve\hat{\bf e}_x+ik_\perp\hat{\bf e}_z)e^{\mp q_\ve z}.
\label{77}
\end{equation}
From this the boundary conditions give the equations (keeping $\mu$ although $\mu=1$ is considered so far)
\begin{eqnarray}
\nonumber
\mu_1+B&=&C+C_1\\
\nonumber
\frac{q_{\ve_1}}{\mu_1}\left(\mu_1-B\right)&=&q( C-C_1)\\
\label{78}
Ce^{-qa}+C_1e^{qa}&=&De^{-q_{\ve_2}a}\\
\nonumber
q(Ce^{-qa}+C_1e^{qa})&=&\frac{q_{\ve_2}}{\mu_2} De^{-q_{\ve_2}a}
\end{eqnarray}
(The coefficients are relative to the $\mu_1$ term from \eqref{66}.) The solution of these equations is
\begin{equation}
D=D_E^{\perp}=\frac{4\mu_1\mu_2\kappa_1 e^{(q_{\ve2}-q)a}}{(\kappa_1+\mu_1)(\kappa_2+\mu_2)[1-B_n e^{-2qa}]},
\label{79a}
\end{equation}
\begin{equation}
B_n=\frac{(\kappa_1-\mu_1)(\kappa_2-\mu_2)}{(\kappa_1+\mu_1)(\kappa_2+\mu_2)}.
\label{79b}
\end{equation}

The Casimir force per unit area between the plates is given by Eq.~(I6.32). This follows from Eq.~(I4.4) where the product of the interaction (\ref{60}) and the Green  function (\ref{69}) multiplied by $-q$ are integrated. With extension to plates with different dielectric constants this becomes
\begin{equation}
f_{sur}=-\frac{1}{\pi\beta}\sum\limits_{n=0}^\infty{^\prime}\int\limits_{\zeta_n}^\infty q^2\,dq \, IQS.
\label{80}
\end{equation}
\begin{equation}
I=\int\limits_0^\infty \int\limits_0^\infty \frac{e^{-(q_{\ve_2}+q)(u_2+a)}
e^{-(q_{\ve_1}+q)u_1}}{q_{\ve_1}q}\,du_1du_2=
\frac{e^{-(q_{\ve_2}+q)a}}{q_{\ve_1}q(q_{\ve_1}+q)(q_{\ve_2}+q)}.
\label{81}
\end{equation}
(with $z=z_2-z_1=u_1+u_2+a$).
The prime means that the $n=0$ term is to be taken with half weight, and $\zeta_n=2\pi n/(\beta\hbar c)$ as follows from (\ref{40}). The $Q$ is the $(9y/2)^2A$ term of (I6.32) which with (I5.5) is $(\ve-1)^2/4$. For different media this becomes
\begin{equation}
Q=(\ve_1-1)(\ve_2-1)/4.
\label{82}
\end{equation}
Further from (I5.5) the $3y=(4\pi/3)\rho\beta\langle s^2\rangle$ which is the classical case. Now the polarizability  $\alpha=\beta\langle s^2\rangle/3$ by which $3y=4\pi\rho\alpha$. This also holds for the quantum case where non-zero frequencies contribute, $\alpha=\alpha(\omega)$. This follows from Sec.~5 of Ref.~\cite{brevik88}.) Finally the last factor is
\begin{equation}
S=S_{EE}=S^{||}+S^\perp, \quad S^{||}=D_E^{||}\langle H_{\ve1}H_1^*\rangle, \quad S^\perp=D_E^{\perp}\langle H_2^2\rangle
\label{83}
\end{equation}
where $H_1=g_1 g_2$, $g_i=iq\hat s_{\perp i}\pm k_\perp \hat s_{||i}$, i.e. $H_1=H_{\ve 1}$ for $\ve=1$. And from the Fourier transform relation \eqref{48a} the $H_1(-k_\perp)$ is the quantity needed. For the complex conjugate we have $H_1^*(k_\perp)=H_1(-k_\perp)$).

With extension of Eqs.~(I6.27) - (I6.30) to different, but non-magnetic, media we now get by use of Eqs.~\eqref{67a} and \eqref{67b} and then Eqs.~\eqref{61} and \eqref{62} ($\langle\hat s_{\perp i}^2\rangle=1/3$ etc.)
\begin{equation}
\langle H_{\ve1}H_1^*\rangle=\frac{1}{9}(k_\perp^2+q_{\ve_1}q)(k_\perp^2+q_{\ve_2}q),
\label{84}
\end{equation}
\begin{equation}
k_\perp^2+q_{\ve_i}q=\frac{1}{\ve_i-1}[\ve_i q^2-q_{\ve_i}+(\ve_i-1)q_{\ve_i}q]=\frac{q}{\ve_ i-1}(\ve_i-\kappa_i)(q+q_{\ve_i}),\quad (i=1,2),
\label{85}
\end{equation}
\begin{equation}
\langle H_2^2\rangle=\frac{1}{9}\zeta^4; \quad \zeta^2=\frac{1}{\ve_ i-1}(q_{\ve_i}^2-q^2)=\frac{q}{\ve_ i-1}(\kappa_i-1)(q_{\ve_i}+q).
\label{86}
\end{equation}
Note these expressions for $k_\perp^2+q_{\ve_i}q$ and $\zeta^2$ are valid only for $\mu_1=\mu_2=1$. Altogether for this case with $I$, $Q$, and $S$ inserted in (\ref{80}) the known Lifshitz result (I2.9) for the force is recovered \cite{lifshitz55}
\begin{equation}
f_{sur}=-\frac{1}{\pi\beta}\sum\limits_{n=0}^\infty{^\prime}\int\limits_{\zeta_n}^\infty q^2\,dq \,\left[\frac{A_n e^{-2qa}}{1-A_ne^{-2qa}}+\frac{B_n e^{-2qa}}{1-B_ne^{-2qa}}\right]
\label{87}
\end{equation}
with $A_n$ and $B_n$ given by Eqs.~(\ref{74b}) and (\ref{79b}) respectively and with $q_{\ve_i}$ ($i=1,2$) given by Eq.~(\ref{62}).

\section{Induced interaction with magnetic properties included}
\label{sec5}

In this section we want to show that expression \eqref{87} for the Casimir force is valid also when magnetic interactions are present, i.e. $\mu_1$ and/or $\mu_2$ are different from one.
In the previous section Sec.~\ref{sec4} the electric field from dipoles and their interactions for a pair of  half-planes were found. This also included the  induced magnetic fields and influence from magnetic permeabilities. Thus the coefficients $D_E^{||}$ and $D_E^\perp$ will stay unchanged. Also the main structure of the surface force (\ref{80}) and integral (\ref{81}) will be the same. However, expression (\ref{83}) will have additional terms that involve magnetic moments too. For these additional terms the $Q$ given by Eq.~(\ref{82}) is modified by replacing $\ve_i$ ($i=1,2$) with  $\mu_i$ in the ways possible. Further the symmetry of electric and magnetic fields in Maxwell's equations (\ref{30}) is utilized. The fully new term to obtain is the induced interaction between electric and magnetic dipole moments.

With $\mu\neq1$ expressions (\ref{83}) and (\ref{84}) for induced interaction between electric dipole moments will remain unchanged, but results (\ref{85}) and (\ref{86}) will not hold any longer. We find the (\ref{85}) can be modified in two useful ways ($q_\ve^2=k_\perp^2+\ve\mu\zeta^2$, $\kappa=q_\ve/q$)
\begin{eqnarray}
\nonumber
k_\perp^2+q_\ve q&=&\frac{1}{\ve-1}[\ve k_\perp^2-k_\perp^2+(\ve-1)q_\ve q]\\
&=&\frac{1}{\ve-1}[(\ve q^2-\ve\zeta^2)-(q_\ve^2-\ve\mu\zeta^2)+(\ve-1)q_\ve q]
\label{90}\\
\nonumber
&=&\frac{q}{\ve-1}[\ve-\kappa)(q+q_\ve)+(\mu-1)\ve\zeta^2],
\end{eqnarray}
and likewise from symmetry with respect to $\ve$ and $\mu$
\begin{equation}
k_\perp^2+q_\ve q=\frac{q}{\mu-1}
[(\mu-\kappa)(q+q_\ve)+(\ve-1)\mu\zeta^2].
\label{91}
\end{equation}

Due to symmetry the contributions from the induced interactions between magnetic dipole moments will be like the ones of Eq.~(\ref{83}) since with $q_\ve^2=k_\perp^2+\ve\mu\zeta^2$ or with \eqref{90} and \eqref{91} the $k_\perp^2+q_\ve q$ stays unchanged by interchange of $\mu$ and $\ve$
\begin{equation}
S_{HH}=D_H^{||}\langle H_{\ve1}H_1^*\rangle+D_H^{\perp}\langle H_2^2\rangle.
\label{92}
\end{equation}
\\tc{red}{Further comparing} with expressions (\ref{74a}) and (\ref{79a}) with interchange of $\mu$ and $\ve$ one finds
\begin{equation}
D_E^\perp=\mu_1\mu_2 D_H^{||}, \quad D_H^\perp=\ve_1 \ve_2 D_E^{||}.
\label{93}
\end{equation}

The remaining terms are those from induced interactions between electric and magnetic dipoles. Thus we need the magnetic fields due to the electric ones and vice versa that follows from symmetry.

With (\ref{66}), (\ref{68}), and (\ref{69}) the TM electric field for $z_2>a$ can be written in the form ${\bf E}=(J_\ve/s_2) D_E^{||}g_{\ve_1 1}(q_{\ve_2}/k_\perp){\bf u}_{ {\ve_2}\pm}$. Thus with Eq.~(\ref{72}) the corresponding magnetic field will be
\begin{equation}
\zeta\mu_2 {\bf H}=\zeta{\bf B}=(J_\ve/s_2) D_E^{||} g_{\ve_1 1}(\pm ik_\perp\ve_2 \mu_2 \zeta^2 \hat {\bf e}_y)/k_\perp
\label{94}
\end{equation}
with $g_{\ve1}$ given by Eq.~(\ref{67b}). This is to be multiplied with the magnetic dipole moment ${\bf s}_2$ (instead of electric dipole moment) whose unit vector has the transverse component $\hat{\bf s}_2^\perp$ ( in the $y$ direction) to obtain the equivalent of $H_{\ve1}$ in Eq.~\eqref{69} in view of the common prefactor $J_\ve D_E^{||}$
\begin{equation}
H_{\ve EH}^{||}=\frac{1}{\zeta\mu_2 J_\ve D_E^{||}}{\bf H}\cdot{\bf s}_2=\pm \zeta g_{\ve_1 1} \ve_2 \hat s_2^\perp.
\label{95}
\end{equation}
This multiplied with the corresponding quantity in vacuum is averaged over the orientations of the dipole moments to give
\begin{equation}
\langle H_{\ve EH}^{||} H_{EH}^{||}\rangle_1=-\frac{1}{9}\zeta^2(k_\perp^2+q_{\ve_1} q)\ve_2.
\label{96}
\end{equation}
Note in \eqref{95} the magnitude of the electric dipole moment $s_2$ is replaced by the magnetic one. This is taken into account by replacing $\ve_2-1$ with $\mu_2-1$ in Eq.~\eqref{82} for the factor $Q$ to be used in Eq.~\eqref{100} below.
The $H_{EH}^{||}=H_{\ve EH}^{||}$ for $\ve_i=\mu_i=1$ ($i=1,2$) with $\zeta\rightarrow -\zeta$. The $\zeta\rightarrow -\zeta$ (and $k_\perp\rightarrow -k_\perp)$ again follows from the Fourier transform relation \eqref{48a}.
Likewise one has the interaction between an electric dipole in halfplane 2 with a magnetic one in halfplane 1. Due to symmetry the corresponding expression must be
\begin{equation}
\langle H_{\ve EH}^{||} H_{EH}^{||}\rangle_2=-\frac{1}{9}\zeta^2(k_\perp^2+q_{\ve_2} q)\ve_1
\label{97}
\end{equation}
Finally for the TM mode one has the interaction between the magnetic moments in the two halfplanes. But again due to symmetry of Maxwell's equations (\ref{30}) this for the magnetic field (in the $y$ direction) is equivalent to the TE field. Thus the corresponding magnetic interactions will be the same and follow from the general form of (\ref{86}) as $\langle H_2^2\rangle=\zeta^2/9$.

The resulting surface force is obtained by adding the various contributions multiplied with factors $Q$ given by (\ref{82}) with one or both $\ve$ replaced by $\mu$. Adding together for the TM contribution the product $QS^{||}$ from Eqs.~(\ref{82}) and \eqref{83} is replaced by
\begin{eqnarray}
\nonumber
S_Q^{||}&=&\frac{1}{4}[(\ve_1-1)(\ve_2-1)D_E^{||}\langle H_{\ve1}H_1^*\rangle+(\ve_1-1)(\mu_2-1)D_E^{||}\langle H_{\ve EH}^{||}H_{EH}^{||}\rangle_1\\
&+&(\ve_2-1)(\mu_1-1)D_E^{||}\langle H_{\ve EH}^{||}H_{EH}^{||}\rangle_2+(\mu_1-1)(\mu_2-1)D_H^{\perp}\langle H_2^2\rangle].
\label{100}
\end{eqnarray}
Inserting from Eqs.~\eqref{84}, \eqref{93}, \eqref{96}  , and \eqref{97} one finds
\begin{eqnarray}
\nonumber
S_Q^{||}&=&\frac{D_E^{||}}{36}[(\ve_1-1)(\ve_2-1)(k_\perp^2+q_{\ve_1}q)(k_\perp^2+q_{\ve_2}q)-(\ve_1-1)(\mu_2-1)\ve_2\zeta^2(k_\perp^2+q_{\ve_1}q)\\
&-&(\ve_2-1)(\mu_1-1)\ve_1\zeta^2(k_\perp^2+q_{\ve_2}q)+(\mu_1-1)(\mu_2-1)\ve_1 \ve_2\zeta^4]
\label{101}\\
&=&\frac{D_E^{||}}{36}[(\ve_1-1)(k_\perp^2+q_{\ve_1}q)-(\mu_1-1)\ve_1\zeta^2][(\ve_2-1)(k_\perp^2+q_{\ve_2}q)-(\mu_2-1)\ve_2\zeta^2].
\nonumber
\end{eqnarray}
Finally with use of Eq.~\eqref{90} the result is
\begin{equation}
S_Q^{||}=\frac{D_E^{||}}{36}q^2(\ve_1-\kappa_1)(q+q_{\ve_1})(\ve_2-\kappa_2)(q+q_{\ve_2}).
\label{102}
\end{equation}
This result is precisely the result for $S_Q^{||}=QS^{||}$ wth Eqs.~\eqref{84} and \eqref{85} inserted. However, the former result was only valid for $\mu=1$ while result \eqref{102} is valid for general $\mu$.

To obtain the resulting force the contribution from the TE mode with magnetic interaction included (i.e. $\mu\neq 1$) is also needed. Again one may find the magnetic interactions via the corresponding magnetic field \eqref{77}. However, we simplify by utilizing the symmetry between the electric and magnetic fields in Maxwells equations. Thus for the magnetic field the TE mode will be similar to the TM mode with exchange of $\ve$ and $\mu$. The various contributions will be similar to those of Eq.~\eqref{100} including the $S^\perp$ of \eqref{83}. Adding together Eq.~\eqref{101} will be replaced by
\begin{equation}
S_Q^\perp=\frac{D_E^\perp}{36\mu_1 \mu_2}[(\mu_1-1)(k_\perp^2+q_{\ve_1}q)-(\ve_1-1)\mu_1\zeta^2][(\mu_2-1)(k_\perp^2+q_{\ve_2}q)-(\ve_2-1)\mu_2\zeta^2].
\label{103}
\end{equation}
With Eqs.~\eqref{91} and\eqref{93} this becomes
\begin{equation}
S_Q^\perp=\frac{D_E^\perp}{36\mu_1\mu_2}q^2(\mu_1-\kappa_1)(q+q_{\ve_1})(\mu_2-\kappa_2)(q+q_{\ve_2}).
\label{104}
\end{equation}
With $\mu_1=1$ and $\mu_2=1$ this result is the same as the result $S_Q^\perp=QS^\perp$ with Eq.~\eqref{86} inserted in \eqref{83}. Result \eqref{104} generalizes expression \eqref{87} for the Casimir force to arbitrary $\mu_1$ and $\mu_2$.

\section{Summary}
\label{sec6}

We have studied the induced Casimir force between media that can have both dielectric and magnetic properties. Usually this force is attractive, but with both properties present it  turns out that it can be repulsive. It is not obvious how this can be understood on physical grounds. In  Sec.~\ref{sec2} we established a simple oscillator model that shows precisely this induced repulsive behavior. The model is a simple extension of the model studied earlier \cite{hoye03,hoye16}, where we  showed the analog of the attractive forces produced by the TM and TE modes of dielectric media. Then by use of the statistical  mechanical method introduced in Ref.~\cite{brevik88} we evaluated the  induced force between a pair of particles possessing  both dielectric and magnetic properties. Further in Sec.~\ref{sec4} the statistical mechanical theory used in Ref.~\cite{hoye98} was extended to a pair of half-planes with different dielectric constants. Finally, in Sec. \ref{sec5} we extended the theory to a pair of half-planes that have both the mentioned  properties and are separated by a distance $a$. The result of the latter, as might be expected, is a generalization of the well known Lifshitz formula as given by Eq.~\eqref{87}. Again the Casimir force is repulsive if of the media is mainly dielectric while the other is mainly magnetic. With equal media in the two half-planes, the Casimir force is always attractive.


\section*{Acknowledgment}
We acknowledge financial support from the Research Council of Norway, Project 250346.

\end{document}